\documentclass[11pt]{article}

\usepackage{latexsym}
\usepackage{amsmath}
\usepackage{color}
 \usepackage{adjustbox}
\usepackage{setspace}

\begin{document}

\newcommand{\qed}{\hfill$\Box$}
\newcommand{\proof}{\noindent {\bf Proof. }}
\newcommand{\eproof}{\hfill $\Box$ \vspace{0.2cm}}

\newtheorem{theorem}{Theorem}
\newtheorem{corollary}{Corollary}
\newtheorem{lemma}{Lemma}
\newtheorem{remark}{Remark}
\newtheorem{example}{Example}
\newtheorem{definition}{Definition}
\newtheorem{notation}{Notation}
\newtheorem{summary}{Summary}
\newtheorem{proposition}{Proposition}

\title{On the $E(s^2)$-optimality of two-level supersaturated designs constructed using Wu's method of partially aliased interactions on certain two-level orthogonal arrays}

\author{
{\protect {
\begin{tabular}{c}
Emmanouil Androulakis$^1$, Kashinath Chatterjee$^2$, Haralambos Evangelaras$^1$\\
$^1$ Department of Statistics and Insurance Science, University of Piraeus, Greece.\\
$^2$ Department of Biostatistics, Data Science and Epidemiology,\\ Augusta University School of Public Health, United States.
\end{tabular}
}}}

\date{}

\maketitle

\begin{abstract}
Wu \cite{Wu} proposed a method for constructing two-level supersaturated designs by using a Hadamard design with $n$ runs and $n-1$ columns as a staring design and by supplementing it with two-column interactions, as long as they are partially aliased. Bulutoglu and Cheng \cite{BulCh} proved that this method results in $E(s^2)$-optimal supersaturated designs when certain interaction columns are selected. In this paper, we extend these results and prove $E(s^2)$-optimality for supersaturated designs that are constructed using Wu's method when the starting design is any orthogonal array with $n$ runs and $n-1$, $n-2$ or $n-3$ columns, as long as its main effects and two-column interactions are partially aliased with two-column interactions.

\smallskip



\end{abstract}

\section{Introduction}
Two-level screening designs have been traditionally exploited to evaluate the main effects of a number of factors (say $f$) using few trials, $n$. The usual model of interest is the first order model  
$$Y = b_0 + \sum_{i=1}^{f} b_i x_i +\epsilon,$$ 
that can be used to evaluate the main effects of the factors using its corresponding $n\times f+1$ model matrix. A two-level design is called supersaturated, if $n<f+1$. Booth and Cox \cite{BC} gave the first systematic construction of supersaturated designs with $n$ runs and $m \ge n$ columns using a simple algorithm and a computer search to find good designs. They restricted their search in the class of balanced designs, so $n$ is even. For the evaluation of their designs, they introduced the $E(s^2)$ criterion which is defined as the average of the squared non-diagonal entries of the matrix $X^T X$, where $X$ is the model matrix of the first order model $$Y = \sum_{i=1}^{m} b_i x_i +\epsilon,$$ without the intercept. A $n\times m$ balanced two-level supersaturated design is $E(s^2)$-optimal if it minimizes $E(s^2)$ among all competitive balanced $n\times m$ two-level designs. Later, several authors dealt with deriving lower bounds of the values of $E(s^2)$ as well as constructing $E(s^2)$-optimal two-level supersaturated designs. For a nice overview we refer to Das et al. \cite{DDCC} and Georgiou \cite{Georgiou} and the references therein.

Wu \cite{Wu} proposed a method for constructing two-level supersaturated designs by using a Hadamard design $H(n,n-1)$ with $n$ runs and $n-1$ columns as a starting design and by supplementing it with two-column interactions, as long as they are partially aliased. Hadamard designs arise from normalized Hadamard matrices when deleting the first column of ones and belong to the class of saturated orthogonal arrays, $OA(n,n-1,2,2)$. However, Wu \cite{Wu} did not give optimality results with respect to the $E(s^2)$ criterion for the supersaturated designs that are constructed using the proposed method. Bulutoglu and Cheng \cite{BulCh} used any Hadamard design when $n \equiv$ 4 (mod 8) or any $n \times n-1$ Paley design when $n \equiv$ 0 (mod 8) as a starting design and proved the $E(s^2)$-optimality of certain supersaturated designs that are constructed using:

\begin{itemize}
\item the starting design $H(n,n-1)$ with all its $n-1 \choose 2$ two factor interactions added.
\item only the $n-1 \choose 2$ two factor interactions of the starting design $H(n,n-1)$.
\item the starting design $H(n,n-1)$ with the two factor interactions of a given column of $H(n,n-1)$ added.
\end{itemize}  

In this paper, we revisit Wu's construction method for two-level supersaturated designs and construct $E(s^2)$-optimal supersaturated designs by employing, as the starting design, any orthogonal array with $n$ runs and $n-1$, $n-2$, or $n-3$ columns, provided that its two-column interactions are partially aliased. Such starting designs can be achieved by removing 0, 1 or 2 columns respectively from any Hadamard design when $n \equiv 4 \pmod{8}$, or from any $n \times (n-1)$ Paley design when $n \equiv 0 \pmod{8}$. Our results rely on the framework of the Generalized Wordlength Pattern (GWP) and its connection to the concept of J-characteristics, offering new insights into the structure and performance of supersaturated designs derived from orthogonal arrays. To deal with the $E(s^2)$-optimality of the constructed designs we follow the notation and the results of Das et al. \cite{DDCC}, who showed that if $X$ is a two-level design with $n \equiv 0$ (mod 4) runs and $m=a(n-1)\pm r$ columns, $a$ positive and $0 \le r \le n/2$, then the average of the squared non-diagonal elements
of $X^TX$ is equal or greater to 
$$\frac{n^2(m-n+1)}{(n-1)(m-1)} +\frac{n}{m(m-1)}(D(n,r)-\frac{r^2}{n-1}),$$ 
where 
$$
D(n,r) = \begin{cases} 4r, & \mbox{if } r\equiv 0 \mbox{ (mod 4)} \\ 
                   n+2r-3, & \mbox{if } r\equiv 1 \mbox{ (mod 4)} \\ 
									   2n-4, & \mbox{if } r\equiv 2 \mbox{ (mod 4)} \\
									 n+2r+1, & \mbox{if } r\equiv 3 \mbox{ (mod 4)} 
					\end{cases}
$$

The rest of the paper is organized as follows: In Section~\ref{prelim} we introduce the required preliminaries, including notation and background on $J$-characteristics relevant to the proposed methodology. 
In Section~\ref{secJ}, we derive closed-form expressions for the sum of squares of the $J$-characteristics, corresponding to specific forms of orthogonal arrays. 
In Section~\ref{mainsec} we investigate the $E(s^2)$-optimality of certain two-level supersaturated designs constructed using Wu’s method, based on the theoretical results established in the preceding Section. Concluding remarks are provided in Section~\ref{conclusion}.

\section{Preliminaries}\label{prelim} 
Let $H(n,q)$ be an orthogonal array $OA(n,q,2,t)$ with $n$ runs and $q$ columns and with a corresponding Generalized Wordlength Pattern (GWP), ${\bf A^g}(H)=$($A^g_1$, $A^g_2$, $A^g_3$, $A^g_4$, \ldots $A^g_q$). Denote with $c_i$, $i=1,2,\ldots q$ its $q$ columns. The values $A^g_1$, $A^g_2$, $\ldots$, $A^g_q$ of the GWP of a two-level array $H(n,q)$ with $n$ runs and $q$ columns are calculated using 
$$A^g_i = \frac{1}{n} \sum_{j=0}^q P_i (j;q) E_j (H) , \; i=1,\ldots,q$$
\noindent
where $P_i (j;q)$ are the Krawtchouk polynomials and $E_j (H), j=0, 1, \ldots, q$ is the distance distribution of $H$, which is given by 
$$ E_j (H) = \frac{\# \{( {\bf c}_1,{\bf c}_2) |, {\bf c}_1, {\bf c}_2 \in H, d_H ({\bf c}_1,{\bf c}_2) =j \}}{n}$$
\noindent
where $d_H ({\bf c}_1,{\bf c}_2)$ is the Hamming distance of two rows ${\bf c}_1$ and ${\bf c}_2$ of $H$. For further details on the aforementioned description, one may refer to Ma and Fang \cite{MF} and Xu and Wu \cite{XW} who applied this procedure to define design aberration criteria. 

The GWP vector ${\bf A^g}(H)$ of $H(n,q)$ can also be calculated by the $J$-characteristics of $H(n,q)$ \cite{DT}, as analyzed by Tang and Deng \cite{TD}. For every {\it s}-subset $S=\{c_{j_1}, c_{j_2},\ldots, c_{j_s}\}$ of columns of $H(n,q)$ with cardinality $|S|=s$, the $J$-characteristics of order $s$ of $H(n,q)$ are calculated by 
$$J_s(S) \equiv J_s(c_{j_1}, c_{j_2},\ldots, c_{j_s})=\sum_{i=1}^n c_{ij_1} \ldots c_{ij_s}.$$
\noindent
Then, for $1 \le s \le q$, all the $J_s (S)$ values of the design $H(n,q)$ with $n$ runs and $q$ columns can be summarized with the value 
$$A_s^g= \frac{1}{n^2} \sum_{|S|=s} J_s(S)^2$$
which corresponds to the $s$ entry of the GWP vector of $H(n,q)$. Therefore,
\begin{equation}\label{main}
\sum_{|S|=s} J_s(S)^2 = n^2 A_s^g.
\end{equation}
This connection has been also given in Xu and Wu \cite{XW} as well as in Bulutoglu and Ryan \cite{BulRy}.

Let us now use a design $H(n,q)$ and follow the method of Wu \cite{Wu} to create a supersaturated design $X$ by augmenting $H$ with the columns that correspond to all two-factor interactions of the columns of $H$, as long as $n \le q + {q\choose 2}$ and there are no fully aliased columns. For this case, the $X^T X$ matrix has the following form

{\tiny
$$\bordermatrix{
          & c_1                 & c_2                 &\ldots&c_q             & c_1c_2                  & c_1c_3        &\ldots & c_{q-1}c_q\cr
c_1       & n                   & J_2(c_1,c_2)        &\ldots&J_2(c_1,c_q)    & J_1(c_2)                & J_1(c_3)        &\ldots &J_3(c_1,c_{q-1},c_q)\cr
c_2       & J_2(c_1,c_2)        & n                   &\ldots&J_2(c_2,c_q)    & J_1(c_1)                & J_3(c_1,c_2,c_3)    &\ldots &J_3(c_2,c_{q-1},c_q)\cr
\vdots    & \vdots              &\vdots               &\ddots&\vdots          & \vdots                  & \vdots        &\ddots &\vdots\cr
c_q       & J_2(c_1,c_q)        & J_2(c_2,c_q)        &\ldots&n               & J_3(c_1,c_2,c_q)        & J_3(c_1,c_3,c_q)    &\ldots &J_1(c_{q-1})\cr
c_1c_2    & J_1(c_2)            & J_1(c_1)            &\ldots&J_3(c_1,c_2,c_q)& n                       & J_2(c_2,c_3)   &\ldots &J_4(c_1,c_2,c_{q-1},c_q)\cr
c_1c_3    & J_1(c_3)            & J_3(c_1,c_2,c_3)    &\ldots&J_3(c_1,c_3,c_q)& J_2(c_2,c_3)            & n             &\ldots &J_4(c_1,c_3,c_{q-1},c_q)\cr
\vdots    & \vdots              &\vdots               &\ddots&\vdots          & \vdots                  & \vdots        &\ddots &\vdots\cr
c_{q-1}c_q& J_3(c_1,c_{q-1},c_q)& J_3(c_2,c_{q-1},c_q)&\ldots&J_1(c_{q-1})    & J_4(c_1,c_2,c_{q-1},c_q)& J_4(c_1,c_3,c_{q-1},c_q)&\ldots &n \cr
 								}$$
}
\noindent and evidently, the $E(s^2)$ value is a function of the $J$-characteristics of $H(n,q)$, up to order $s=4$. In this matrix every $J_3(S)$ value appears 6 times in total since, for example, $J_3(c_1, c_2, c_3)$ appears in cells ($c_1$, $c_2c_3$), ($c_2$, $c_1c_3$), ($c_3$, $c_1c_2$), ($c_2c_3$, $c_1$), ($c_1c_3$, $c_2$) and ($c_1c_2$, $c_3$). Similarly,  every $J_4(S)$ value appears 6 times in total. Finally, all $J_1(S)$ and $J_2(S)$ values are equal to zero, since $D(H,q)$ is an orthogonal array. Consequently, in order to evaluate a supersaturated design in terms of the $E(s^2)$, it is essential to quantify the contribution of each $J$-characteristic term. This motivates a detailed examination of the sum of squares of the $J$-characteristics, which forms the focus of the next Section.

\section{Useful results on $J$-characteristics} \label{secJ}

In this Section, we calculate the sum of squares of the $J$-characteristics for an orthogonal array $H(n, q)$, where $q = n - 1$, $n - 2$, $n - 3$, and $n - 4$, using equation \ref{main}. The key results are summarized in the following Lemmas.
Specifically, Lemma \ref{lemma1} presents the results for the case where all columns of $H(n, q)$ are included in the sets $S$ of cardinality 3 and 4. Lemma \ref{lemma1a} provides the corresponding sum of squares when only certain columns are considered in $S$.

\begin{lemma}\label{lemma1}
Let $H(n,n-1)$ be a saturated orthogonal array $OA(n,n-1,2,2)$ and let $H(n,n-2)$, $H(n,n-3)$ and $H(n,n-4)$ be the corresponding orthogonal arrays with $n$ runs and $n-2$, $n-3$ and $n-4$ columns, respectively, that are produced from $H(n,n-1)$ after deleting one, two and three columns. Let also denote with $J_{3,q}(c_i, c_j, c_k)$ and $J_{4,q}(c_i, c_j, c_k, c_l)$, $i<j<k<l$, the $J$-characteristics of $H(n,q)$, $q=n-1, n-2, n-3$ and $n-4$ as calculated using any three and any four distinct columns of $H(n,q)$, respectively. Then, 

\[
\begin{array}{ll}
1.& \displaystyle \sum J_{3,n-1}^2(c_i, c_j, c_k) = n^2 (n-1)(n-2)/6.\\
2.& \displaystyle \sum J_{3,n-2}^2(c_i, c_j, c_k) = n^2 (n-2)(n-4)/6.\\ 
3.& \displaystyle \sum J_{3,n-3}^2(c_i, c_j, c_k) = n^2 (n-4)(n-5)/6.\\
4.& \displaystyle \sum J_{3,n-4}^2(c_i, c_j, c_k) = n^2 (n-4)(n-8)/6 +16d(n-4d).\\ 
5.& \displaystyle \sum J_{4,n-1}^2(c_i,c_j,c_k,c_l) = n^2 (n-1)(n-2)(n-4)/24.\\
6.& \displaystyle \sum J_{4,n-2}^2(c_i,c_j,c_k,c_l) = n^2 (n-2)(n-4)(n-5)/24.\\
7.& \displaystyle \sum J_{4,n-3}^2(c_i,c_j,c_k,c_l) = n^2 (n-4)(n-5)(n-6)/24.\\
8.& \displaystyle \sum J_{4,n-4}^2(c_i,c_j,c_k,c_l) = n^2 (n-4)(n^2-15n+62)/24 - 16d(n-4d).\\
\end{array}
\]
where $d$ is an non negative integer, $d \le n/4$.
\end{lemma}

\proof
Let $H_n$ be a normalized Hadamard matrix of order $n$. The non-zero elements of its distance distribution are $E_0=1$ and $E_{n/2}=n-1$. If we remove the first column of ones, we obtain a design $H(n,n-1)$ which is actually a saturated $OA(n, n-1, 2, 2)$ with the same distance distribution of $H_n$, since we removed an identical element from each row of $H_n$. Therefore, $A_3 = P_3(0;n-1) E_0 + P_3(n/2;n-1) E_{n/2}$ and $A_4 = P_4(0;n-1) E_0 + P_4(n/2;n-1) E_{n/2}$. The result follows by taking into account that $P_3(0;n-1)=(n-3)(n-2)(n-1)/6$, $P_3(n/2;n-1)=(n-2)/2$, $P_4(0;n-1)=(n-4)(n-3)(n-2)(n-1)/24$ and $P_4(n/2;n-1)=(n-4)(n-2)/8$.

Lets now remove a column from $H(n,n-1)$, to obtain a $H(n,n-2)$. The latter is an $OA(n,n-2,2,2)$ and the non-zero elements of its distance distribution can be easily calculated to be $E_0=1$, $E_{(n-2)/2}=n/2$ and $E_{n/2}=(n-2)/2$, since we removed a balanced column from $H(n,n-1)$. The result is evident since $P_3(0;n-2)=(n-4)(n-3)(n-2)/6$, $P_3((n-2)/2;n-2)=0$, $P_3(n/2;n-2)=(n-4)$, $P_4(0;n-2)=(n-5)(n-4)(n-3)(n-2)/24$, $P_4((n-2)/2;n-2)=(n-4)(n-2)/8$ and $P_4(n/2;n-2)=(n-10)(n-4)/8$.

If we remove two columns from $H(n,n-1)$ we obtain an $OA(n,n-3,2,2)$ and the non-zero elements of its distance distribution are $E_0=1$, $E_{(n-4)/2}=n/4$, $E_{(n-2)/2}=n/2$ and $E_{n/2}=(n-4)/4$ since the pair of columns that were removed constitute $n/4$ replicates of the full $2^2$ design. For this case, $P_3(0;n-3)=(n-5)(n-4)(n-3)/6$, $P_3((n-4)/2;n-3)=(4-n)/2$, $P_3((n-2)/2;n-3)=(n-4)/2$, $P_3(n/2;n-2)=3(n-20)/2$, $P_4(0;n-3)=(n-6)(n-5)(n-4)(n-3)/24$, $P_4((n-4)/2;n-3)=(n-6)(n-4)/8$, $P_4((n-2)/2;n-3)=(n-6)(n-4)/8$ and $P_4(n/2;n-3)=(n-20)(n-6)/8$.

Finally, if we remove three columns from $H(n,n-1)$ we obtain an $OA(n,n-4,2,2)$ and the non-zero elements of its distance distribution can be calculated by exploiting that the three columns that are removed from the $OA(n,n-1,2,2)$ form a design that consists of $d$ replicates of the $2^{3-1}$ fraction with $I=ABC$ and $n/4-d$ replicates of the $2^{3-1}$ fraction with $I=-ABC$, $0 \le d \le n/4$. So, $E_0=1$, $E_{(n-6)/2}=2d(n-4d)/n$, $E_{(n-4)/2}=(96d^2-24dn+3n^2)/4n$, $E_{(n-2)/2}=6d(n-4d)/n$ and $E_{n/2}=(8d(4d-n)+n(n-4))/4n$. The values of the Krawtchouk polynomials are: $P_3(0;n-4)=(n-6)(n-5)(n-4)/6$, $P_3((n-6)/2;n-4)=(6-n)$, $P_3((n-4)/2;n-4)=0$, $P_3((n-2)/2;n-4)=(n-6)$, $P_3(n/2;n-4)=2(n-10)$, $P_4(0;n-4)=(n-7)(n-6)(n-5)(n-4)/24$, $P_4((n-6)/2;n-4)=(n-6)(n-12)/8$, $P_4((n-4)/2;n-4)=(n-6)(n-4)/8$, $P_4((n-2)/2;n-4)=(n-6)(n-12)/8$ and $P_4(n/2;n-4)=(n^2-42n+280)/8$. 
\qed

\begin{lemma}\label{lemma1a}
Under the statements of Lemma \ref{lemma1}, if $c_{i_0}$ and $c_{j_0}$ are two specific columns of $H(n,q)$,
 $q=n-1, n-2, n-3$ and $n-4$, and $d$ is an non negative integer, $d \le n/4$, then 
 
\[
\begin{array}{ll}
1.& \displaystyle \sum J_{3,n-1}^2(c_{i_0}, c_j, c_k) = n^2 (n-2)/2.\\
2.& \displaystyle \sum J_{3,n-2}^2(c_{i_0}, c_j, c_k) = n^2 (n-4)/2.\\
3.& \displaystyle \sum J_{3,n-3}^2(c_{i_0}, c_j, c_k) = n^2 (n-4)/2 -16d(n-4d).\\
4.& \displaystyle \sum J_{3,n-1}^2(c_{i_0}, c_{j_0}, c_k) = n^2.\\
5.& \displaystyle \sum J_{3,n-2}^2(c_{i_0}, c_{j_0}, c_k) = 16d(n-4d).\\
6.& \displaystyle \sum J_{4,n-1}^2(c_{i_0}, c_j, c_k, c_l) = n^2 (n-2)(n-4)/6.\\
7.& \displaystyle \sum J_{4,n-2}^2(c_{i_0}, c_j, c_k, c_l) = n^2 (n-4)(n-5)/6.\\
8.& \displaystyle \sum J_{4,n-3}^2(c_{i_0}, c_j, c_k, c_l) = n^2 (n-4)(n-8)/6 + 16d(n-4d).\\
9.& \displaystyle \sum J_{4,n-1}^2(c_{i_0}, c_{j_0}, c_k, c_l) = n^2(n-4)/2.\\
10.& \displaystyle \sum J_{4,n-2}^2(c_{i_0}, c_{j_0}, c_k, c_l) = n^2 (n-4)/2 -16d(n-4d).
\end{array}
\]
\end{lemma}

\proof 
These results can be easily obtained using Lemma \ref{lemma1} and the following equations  
\[
\begin{array}{l}
\displaystyle \sum J_{3,q}^2(c_i, c_j, c_k) = \sum J_{3,q-1}^2(c_i, c_j, c_k) + \sum J_{3,q}^2(c_{i_0}, c_j, c_k)\\ 
\displaystyle \sum J_{3,q}^2(c_i, c_j, c_k) = \sum J_{3,q-1}^2(c_i, c_j, c_k) + \sum J_{3,q}^2(c_{i_0}, c_{j_0}, c_k) + \sum J_{3,q-1}^2(c_{i_0}, c_j, c_k)\\ 
\displaystyle \sum J_{4,q}^2(c_i, c_j, c_k, c_l) = \sum J_{4,q-1}^2(c_i, c_j, c_k, c_l) + \sum J_{4,q}^2(c_{i_0}, c_j, c_k, c_l)\\ 
\displaystyle \sum J_{4,q}^2(c_i, c_j, c_k, c_l) = \sum J_{4,q-1}^2(c_i, c_j, c_k, c_l) + \sum J_{4,q}^2(c_{i_0}, c_{j_0}, c_k, c_l) + \sum J_{4,q-1}^2(c_{i_0}, c_j, c_k, c_l)\\ 
\end{array}
\]
\noindent
that arise from the relations that the $J$-characteristics of $H(n,q)$, $q=n-1$, $n-2$ and $n-3$ possess.
\qed

\vspace{1em}
These Lemmas can be used to enhance the calculation of the $E(s^2)$ value of a supersaturated design that is constructed by extending an $OA(n,q,2,2)$, $q=n-1$, $n-2$ and $n-3$ with interaction columns. In the following Section we prove the $E(s^2)$-optimality of certain supersaturated designs.

\section{$E(s^2)$-optimality of certain supersaturated designs constructed using Wu's method}\label{mainsec}

Using the results derived in Section \ref{secJ}, we now establish the $E(s^2)$-optimality of certain supersaturated designs constructed via Wu's method \cite{Wu} of partially aliased interactions, provided that the starting array comes from a Hadamard design when $n \equiv 4 \pmod{8}$, or a $n \times (n-1)$ Paley design when $n \equiv 0 \pmod{8}$ after removing 0, 1 or 2 columns as explained in the Introduction. Theoretical justification and supporting results are provided in the following Theorems.

\begin{theorem}\label{th1}
Let $H(n,q)$ be an orthogonal array $OA(n,q,2,2)$, $q=n-1$, $n-2$ or $n-3$. The $n \times q(q+1)/2$ design $X$, which is produced by augmenting $H(n,q)$ with its two-column interactions, is an $E(s^2)$-optimal two-level supersaturated design.
\end{theorem}

\proof
Following the notation and results of Das et al. \cite{DDCC}, we obtain:

 	\begin{itemize}
	\item  Case 1: $q=n-1$. We have that $m = \frac{n}{2} (n-1),$ so $a =  \frac{n}{2}$ and $r =0 \equiv  0$ mod 4. The lower bound of Das et al. \cite{DDCC}, denoted as $LB$, will be $$LB = \frac{n^2}{n+1}.$$ 
	 Let $s_{ij}$, $1\leq i,j\leq q(q+1)/2$, $i\neq j$, the $(i,j)$-th entry of $X^TX$. The average of squares of its non-diagonal elements, denoted as $E(s^2)$, will be
	 $$
	 E(s^2) = \frac{6 \sum J_{3,n-1}^2(c_i, c_j, c_k) + 6\sum J_{4,n-1}^2(c_i,c_j,c_k,c_l)}{m(m-1)},
	 $$
	 taking into account the number of possible appearances of the corresponding non-zero  $J$-characteristics in $X^TX$.
	 Based on the statements of Lemma 1, we get that $$E(s^2) = \frac{n^2(n-1)(n-2) + \frac{n^2(n-1)(n-2)(n-4)}{4}}{\frac{n}{2}(n-1) \big(\frac{n}{2}(n-1)-1\big)} = \frac{n^2}{n+1}.$$

\item  Case 2: $q=n-2$. We have that $m = \frac{(n-2)(n-1)}{2}$ so $a = \frac{(n-2)}{2}$ and $r =0 \equiv  0$ mod 4. Therefore $$LB = \frac{n(n-4)}{n-3}.$$ In a similar manner, $E(s^2)$ is calculated as
	 $$
	 E(s^2) = \frac{6 \sum J_{3,n-2}^2(c_i, c_j, c_k) + 6\sum J_{4,n-2}^2(c_i,c_j,c_k,c_l)}{m(m-1)} $$
	 
	 $$ =  \frac{n^2(n-2)(n-4) + \frac{n^2(n-2)(n-4)(n-5)}{4}}{\frac{n-2}{2}(n-1) \big(\frac{n-2}{2}(n-1)-1\big)} =  \frac{n(n-4)}{n-3}.
	 $$

\item  Case 3: $q=n-3$. Therefore, $m=\frac{(n-3)(n-2)}{2} = \frac{(n-4)(n-1)}{2} +1$, so $a = \frac{(n-4)}{2}$ and $r=1 \equiv  1$ mod 4. Consequently, $$LB = \frac{n^2(n-5)}{(n-3)(n-1)}$$ and $E(s^2)$ is calculated as
	 $$
	 E(s^2) = \frac{6 \sum J_{3,n-3}^2(c_i, c_j, c_k) + 6\sum J_{4,n-3}^2(c_i,c_j,c_k,c_l)}{m(m-1)} 
	 $$
	 $$ =  \frac{n^2(n-4)(n-5) + \frac{n^2(n-4)(n-5)(n-6)}{4}}{\bigg(\frac{(n-4)(n-1)}{2} +1\bigg) \bigg(\frac{(n-4)(n-1)}{2}\bigg)} =  \frac{n^2(n-5)}{(n-3)(n-1)}.
	 $$
	 \qed
 
\end{itemize}

\noindent We note again that the case where $q=n-1$ has been also proven in Bulutoglu and Cheng \cite{BulCh}.

\begin{theorem}\label{th2}
	Let $H(n,q)$ be an orthogonal array $OA(n,q,2,2)$, $q=n-1$ or $n-2$. The $n \times \{q(q+1)/2 -1\}$ design $X$, which is produced by augmenting $D(n,q)$ with its two-column interactions, minus one column, is an $E(s^2)$-optimal two-level supersaturated design.
\end{theorem}

\proof
We now must eliminate the $J$-characteristics corresponding to the triplets and/or quadruplets that would be formed by the deleted column. Note that the latter can be either a main effect $c_{i_0}$ or an interaction effect $c_{i_0}c_{j_0}$. Therefore we have the following results:

	\begin{itemize}
	\item Case 1: $q=n-1$. We have that $m = \frac{n}{2} (n-1) -1,$ so $a =  \frac{n}{2}$, $r =1 \equiv  1$ mod 4 and the lower bound of Das et al. \cite{DDCC} becomes $$LB = \frac{n^2}{n+1}.$$
	
	\begin{itemize}
		\item If the deleted column is a main effect $c_{i_0}$, then the average of the squared non-diagonal elements of $X^TX$ will be modified as
	$$
	E(s^2) = \frac{6 \sum J_{3,n-1}^2(c_i, c_j, c_k) + 6\sum J_{4,n-1}^2(c_i,c_j,c_k,c_l) - 2 \sum J_{3,n-1}^2(c_{i_0}, c_j, c_k)}{m(m-1)}.
	$$
	Under the statements of Lemma 1 and Lemma 2, we have that
	$$
	E(s^2) = \frac{n^2(n-1)(n-2) + \frac{n^2(n-1)(n-2)(n-4)}{4} - n^2(n-2)}{\big(\frac{n}{2}(n-1) - 1\big) \big(\frac{n}{2}(n-1) - 2\big)} = \frac{n^2}{n+1}.
	$$
	
	\item  If the deleted column is an interaction effect $c_{i_0}c_{j_0}$, then $E(s^2)$ will be modified as
	$$
	\textstyle Es^2) =  \frac{6 \sum J_{3,n-1}^2(c_i, c_j, c_k) + 6\sum J_{4,n-1}^2(c_i,c_j,c_k,c_l) - 2 \sum J_{3,n-1}^2(c_{i_0}, c_{j_0}, c_k) - 2 \sum J_{4,n-1}^2(c_{i_0}, c_{j_0}, c_k, c_l)}{m(m-1)}
	$$
	$$
	= \frac{n^2(n-1)(n-2) + \frac{n^2(n-1)(n-2)(n-4)}{4} - 2n^2 - n^2(n-4)}{\big(\frac{n}{2}(n-1) - 1\big) \big(\frac{n}{2}(n-1) - 2\big)}  = \frac{n^2}{n+1}.
	$$
	
\end{itemize}
\end{itemize}
	
	\begin{itemize}
	\item Case 2: $q=n-2$. We have that $m = \frac{(n-2)(n-1)}{2} -1,$ so $a =  \frac{n-2}{2}$, $r =1 \equiv  1$ mod 4 and $$LB = \frac{n(n-4)}{n-3}.$$ 
		\begin{itemize}
		\item  Similarly, if the deleted column is a main effect $c_{i_0}$, then $E(s^2)$ will be modified as
		$$
		E(s^2) = \frac{6 \sum J_{3,n-2}^2(c_i, c_j, c_k) + 6\sum J_{4,n-2}^2(c_i,c_j,c_k,c_l) - 2 \sum J_{3,n-2}^2(c_{i_0}, c_j, c_k)}{m(m-1)}
		$$
		$$
		=  \frac{n^2(n-2)(n-4) + \frac{n^2(n-2)(n-4)(n-5)}{4} - n^2(n-4)}{\bigg(\frac{(n-2)(n-1)}{2} -1\bigg) \bigg(\frac{(n-2)(n-1)}{2} -2\bigg)}  = \frac{n(n-4)}{n-3}.
		$$

		\item  If the deleted column is an interaction effect $c_{i_0}c_{j_0}$, then $E(s^2)$  will be
 	$$
 \textstyle E(s^2) =  \frac{6 \sum J_{3,n-2}^2(c_i, c_j, c_k) + 6\sum J_{4,n-2}^2(c_i,c_j,c_k,c_l) - 2 \sum J_{3,n-2}^2(c_{i_0}, {j_0}, c_k) - 2 \sum J_{4,n-2}^2(c_{i_0}, c_{j_0}, c_k, c_l)}{m(m-1)}
 $$
 $$
 = \frac{n^2(n-2)(n-4) + \frac{n^2(n-2)(n-4)(n-5)}{4} - 32d(n-4d) - 2 \big(n^2 (n-4)/2 -16d(n-4d)\big)   }{\bigg(\frac{(n-2)(n-1)}{2} -1\bigg) \bigg(\frac{(n-2)(n-1)}{2} -2\bigg)}  
 $$
 $$ =  \frac{n(n-4)}{n-3}.$$

\end{itemize}
\end{itemize}
 \qed

\begin{remark}
The result of Theorem \ref{th2} can also be obtained from Cheng \cite{cheng} since from Theorem \ref{th1} and for the cases 1 and 2, the columns of the $E(s^2)$-optimal supersaturated design is of the form $a(n-1)$. Using the results of Cheng \cite{cheng} we can also remove an additional column to obtain an $E(s^2)$-optimal supersaturated design.
\end{remark}

\begin{theorem}\label{th3}
	Let $H(n,q)$ be an orthogonal array $OA(n,q,2,2)$, $q=n-1$, $n-2$ or $n-3$. The $n \times q(q-1)/2$ design $X$, which is produced by considering only the two-column interactions of $D(n,q)$, is an $E(s^2)$-optimal two-level supersaturated design.
\end{theorem}

\proof

\noindent The non-zero $J-$characteristics appearing as the non-diagonal elements of $X^TX,$ correspond only to quadruplets of the form $(c_i, c_j, c_k, c_l)$. Therefore, we have the following results:

	\begin{itemize}
	\item  Case 1: $q=n-1$. We have that $m = \frac{(n-2)}{2}(n-1),$ so $a =  \frac{n-2}{2}$, $r =0 \equiv  0$ mod 4 and the lower bound of Das et al. \cite{DDCC} will be $$LB = \frac{n(n-4)}{n-3}.$$ The average of the squared non-diagonal elements of $X^TX$ becomes
$$
E(s^2) = \frac{6 \sum J_{4,n-1}^2(c_i, c_j, c_k, c_l)}{m(m-1)} =  \frac{\frac{n^2(n-1)(n-2)(n-4)}{4}}{ \bigg( \frac{(n-2)(n-1)}{2} \bigg) \bigg( \frac{(n-2)(n-1)}{2}  -1 \bigg) } = \frac{n(n-4)}{n-3}.
$$

\item  Case 2: $q=n-2$. We have that $m = \frac{(n-2)(n-3)}{2} = \frac{(n-4)}{2}(n-1) +1,$ so $a =  \frac{n-4}{2}$, $r =1 \equiv  1$ mod 4 and $$LB = \frac{n^2(n-5)}{(n-1)(n-3)}.$$ In addition, 
$$
E(s^2) = \frac{6 \sum J_{4,n-2}^2(c_i, c_j, c_k, c_l)}{m(m-1)} =  \frac{\frac{n^2(n-2)(n-4)(n-5)}{4}}{ \bigg( \frac{(n-4)(n-1)}{2} +1 \bigg) \bigg( \frac{(n-4)(n-1)}{2} \bigg) } = \frac{n^2(n-5)}{(n-1)(n-3)}.
$$

\item   Case 3: $q=n-3$. We have that $m =  \frac{(n-3)(n-4)}{2} =  \frac{(n-6)}{2} (n-1) +3,$ so $a =  \frac{n-6}{2}$, $r =3 \equiv  3$ mod 4 and $$LB = \frac{ n(n^3 - 13n^2 + 48n - 32) }{(n-3)(n-4)(n-5) }.$$ On the other hand, 
$$
E(s^2) = \frac{6 \sum J_{4,n-3}^2(c_i, c_j, c_k, c_l)}{m(m-1)} =  \frac{\frac{n^2(n-4)(n-5)(n-6)}{4}}{ \bigg( \frac{(n-6)(n-1)}{2} +3 \bigg) \bigg( \frac{(n-6)(n-1)}{2} +2 \bigg) } = \frac{n^2(n-6)}{(n-2)(n-3)}.
$$

Subtracting $LB$ from $E(s^2)$ yields
$$
E(s^2) - LB = 
 \frac{8n\left(n-8\right)}{\left(n-2\right)\left(n-3\right)\left(n-4\right)\left(n-5\right)} \to 0 \quad \text{as} \quad n \to \infty. $$
 
\qed
\end{itemize}

\begin{remark}
With respect to the results of Theorem \ref{th3} we note the following:
	\begin{itemize}
		\item The result of Theorem \ref{th3}, case 1 ($q=n-1$), has also been proven in Bulutoglu and Cheng \cite{BulCh}. Since the columns of the $E(s^2)$-optimal supersaturated design is of the form $a(n-1)$, we can also refer to Cheng \cite{cheng} and remove one or two columns to obtain an $E(s^2)$-optimal supersaturated design.
		\item The case for $q=n-3$ results in designs that are asymptotically $E(s^2)$-optimal. 
	\end{itemize}

\end{remark}

Bulutoglu and Cheng \cite{BulCh} also established the $E(s^2)$-optimality of the case presented in the following Theorem. Here, we provide an alternative proof that is based on our proposed methodology. In addition, we demonstrate  that this construction does not result in $E(s^2)$-optimal supersaturated designs when $q=n-2$ and $q=n-3$.

\begin{theorem}\label{th4}
Let $H(n,q)$ be an orthogonal array $OA(n,q,2,2)$, $q=n-1$. The $n \times \{2q-1\}$ design $X$, which is produced by augmenting $D(n,q)$ with the two-column interactions of one parent factor only, is an $E(s^2)$-optimal two-level supersaturated design. When $q=n-2$ and $q=n-3$ the resulting design is not $E(s^2)$-optimal.
\end{theorem}

\proof

\noindent The non-zero $J-$characteristics appearing as the non-diagonal elements of $X^TX,$ correspond only to the triplets of the form $(c_{i_0}, c_j, c_k)$, created by the parent factor $c_{i_0}$. Therefore, we have the following results:

\begin{itemize}
	\item  Case 1: $q=n-1$. We have that $m = 2(n-1) -1$ so $a = 2$, $r = 1 \equiv  1$ mod 4 and the lower bound of Das et al. \cite{DDCC} will be $$LB = \frac{n^2}{2n-3}.$$ The average of the squared non-diagonal elements of $X^TX$ becomes
	$$
	E(s^2) = \frac{4 \sum J_{3,n-1}^2(c_{i_0}, c_j, c_k)}{m(m-1)} =  \frac{2n^2(n-2) }{\big(2(n-1)-1\big)\big(2(n-1)-2\big)} = \frac{n^2}{2n-3}.
	$$

\qed

	\item Case 2: $q=n-2$. We have that $m = 2(n-2) -1 = 2(n-1)-3$ so $a = 2$, $r = 3 \equiv  3$ mod 4 and $$LB = \frac{n(n^2-5n+8)}{(2n-5)(n-3)}.$$ Moreover, 
	$$
	E(s^2) = \frac{4 \sum J_{3,n-2}^2(c_{i_0}, c_j, c_k)}{m(m-1)} =  \frac{2n^2(n-4) }{\big(2(n-1)-3\big)\big(2(n-1)-4\big)} = \frac{n^2(n-4)}{(2n-5)(n-3)}.
	$$
	Hence, we have that 
	$$
	E(s^2) - LB = \frac{n^2-8n}{\left(2n-5\right)\left(n-3\right)}
	\to \frac{1}{2} \quad \text{as} \quad n \to \infty. $$ 
	Therefore, the resulting design is not optimal.

	\item  Case 3: $q=n-3$. We have that $m = 2(n-3) -1 = 2(n-1)-5$ so $a = 2$,  $r = 5 \equiv  1$ mod 4 and $$LB = \frac{n(n-4)}{2n-7}.$$ The average of the squared non-diagonal elements of $X^TX$ becomes
	$$
	E(s^2) = \frac{4 \sum J_{3,n-3}^2(c_{i_0}, c_j, c_k)}{m(m-1)} =  \frac{ 4\big(n^2 (n-4)/2 -16d(n-4d)\big) }{\big(2(n-1)-5\big)\big(2(n-1)-6\big)} =  \frac{n^3-4n^2-32nd+128d^2}{\left(2n-7\right)\left(n-4\right)}.
	$$	
	Then,
	$$
	E(s^2) - LB = \frac{4n^2+128d^2-32nd-16n}{\left(n-4\right)\left(2n-7\right)} \to 2  \quad \text{as} \quad n \to \infty.  
	$$
	Hence, the resulting design is not optimal.
	
\end{itemize}
		\qed

\section{Conclusion}\label{conclusion}

In this paper, we built upon the framework proposed by Wu~\cite{Wu} for constructing two-level supersaturated designs through partially aliased interactions. Extending the results of Bulutoglu and Cheng~\cite{BulCh}, we demonstrated that supersaturated designs derived from specific classes of orthogonal arrays whose interaction columns are not fully aliased attain $E(s^2)$-optimality. A key element of our methodology is the use of the concept of $J$-characteristics, which provides a unifying framework for quantifying the structural properties of the examined designs and for assessing the contribution of main effects and interaction terms to the $E(s^2)$ criterion. By deriving closed-form expressions for the sum of squares of the $J$-characteristics, we obtained explicit formulas for $E(s^2)$, enabling a rigorous analytical assessment of optimality.

In particular, we have established that designs constructed from the aforementioned class of orthogonal arrays with $q = n - 1$, $n - 2$, or $n - 3$ columns are $E(s^2)$-optimal when all two-factor interaction columns are included. We further investigated several variations of Wu’s method, involving constructions in which a single column is omitted, only interaction terms are employed, or interactions are restricted to those involving a fixed parent factor. For each of these constructions, we analytically evaluated whether the resulting design attains the theoretical lower bound for $E(s^2)$, thereby identifying both optimal and non-optimal cases. Overall, our findings extend the class of known $E(s^2)$-optimal supersaturated designs and demonstrate the utility of the $J$-characteristics framework as a rigorous and effective tool for their theoretical evaluation. Finally, we note that the proposed methodology can also be applied to examine $E(s^2)$-optimality for starting designs with fewer than $n - 3$ columns; however, additional results concerning the corresponding $J$-characteristics, analogous to Lemmas \ref{lemma1} and \ref{lemma1a}, must first be established.

\end{document}